\documentclass[prl,aps,showpacs,amsmath,floatfix,twocolumn]{revtex4}
\usepackage[]{graphicx}
\usepackage{graphicx}
\usepackage{dcolumn}
\usepackage{bm}

\def\bH { { \bf H }}

\def \bea { \begin{eqnarray} }
\def \be { \begin{equation} }
\def \ee { \end{equation} }
\def \eea { \end{eqnarray} }

\def\nlab#1{\label{#1} }
\def\lbl#1{ \nlab{#1}\makebox[.5cm][c]{  { \tiny \bf  #1 }}  }
\def\lbl#1{ \nlab{#1} }

\begin{document}

\title{Strong correlation effects in diatomic
molecular electronic devices}

\author{A. Goker$^{1,2}$, F. Goyer$^1$  and M. Ernzerhof$^1$}
\affiliation{$^1$
Department de Chimie, Universite de Montreal, \\
C.P.6128 Succursale A, Montreal, Quebec H3C 3J7, Canada
}
\affiliation{$^2$
Department of Physics, Fatih University, \\
Buyukcekmece, Istanbul 34500, Turkey
}

\date{\today}

\begin{abstract}
We present a qualitative model for a fundamental process in molecular electronics: 
the change in conductance upon bond breaking. In our model a diatomic molecule 
is attached to spin-polarized contacts. Employing a Hubbard Hamiltonian, electron 
interaction is neglected in the contacts and explicitly considered in the molecule, 
enabling us to study the impact of electron interaction on the molecular conductance. 
In the limit where the electron repulsion is strong compared to the binding energy 
(as it becomes the case upon dissociation) electron transmission in strongly suppressed 
compared to the non-interacting case. However, the spin-polarized nature of the contacts 
introduces a coupling between the molecular singlet and triplet states. This coupling 
in turn yields additional resonances in the transmission probability that significantly 
reduce the energetic separation appear between resonances. The ramifications of our 
results on transport experiments performed on nanowires with inclusions $H_2$ are discussed.
\end{abstract}

\maketitle

Molecular electronics \cite{JGA00,N01,NR03,HR03,CHH07} offers new perspectives
for the construction of electronic devices. Even though some quantum effects,
such as tunneling, are sometimes undesirable in conventional electronics, molecular
electronic devices (MEDs) systematically exploit quantum effects and are therefore
suitable for the construction of devices on the atomic scale. Another field that may
contribute to the development of new electronic devices is spintronics (for reviews see, 
e.g., \cite{Zuticetal04RMP,Bratkovsky08RPP}), where spin and charge transport are 
investigated on a mesoscopic scale. Molecular spintronics emerged in experimental 
and theoretical studies as a result of recent combination of molecular electronics 
and spintronics \cite{Rochaetal05,Seneoretal07JPCM}. Most theoretical 
approaches to MED's are based on effective one-electron theories
\cite{MKR94,D95,NDL95,XDR01,TGW01a,BMO02,KBY04,EZ05,SGP06} and do not 
properly account for electron correlation effects \cite{EZ05,kbe06}. Providing 
a remedy for this shortcoming remains a formidable challenge. Recent work to address 
this issue invokes, for instance, the configuration interaction method to calculate 
the conductance.  While the proper inclusion of the 
infinite contacts remains an unsolved problem in this approach, correlation effects 
are accounted for. Non-equilibrium Green's function techniques have also been employed 
\cite{T08} in conjunction with model Hamiltonians to understand correlation effects. 
Here, we follow a similar route. We use the recently developed source-sink potential 
method (SSP) \cite{GEZ07,Ernzerhof07} to a model Hamiltonian involving Hubbard 
term to address the problem of electron correlation in molecular conductors.

One of the most fundamental phenomena undergone by molecules is that of bond formation
and bond breaking. Chemical bonds are sometimes stretched and broken by mechanical
force in break junction experiments \cite{ME85,RAP96}. In this letter, we address the 
breaking of a chemical bond in a diatomic molecule, which constitutes part of a MED. 
We provide an answer to the question of how the conductance of a diatomic molecule changes 
during bond breaking. In a simple molecular orbital picture (appropriate for non-interacting 
electrons), during the breakage of a bond, the binding and antibinding orbital approach 
each other in energy until they become degenerate in the separated atom limit. Therefore, 
electron interaction effects become increasingly important on the scale of the 
single-electron energies and a strongly correlated system is obtained. This system
cannot be properly described by the Kohn-Sham reference determinant  \cite{PY89,DG90},
or any other single-particle description, appropriate for a non-degenerate ground state. 
As a consequence, the popular combination of DFT and non-equilibrium Green's function technique 
\cite{KMR94b,D95,Tayloretal01PRB}  to model transport in MEDs is inadequate.

The system under consideration is comprised of two semi-infinite non-interacting contacts. 
The boundary conditions of the problem are chosen such that in the left contact (L) we have an 
incoming Bloch wave that provides an electron current towards the contact-molecule interface. 
At the interface, the incoming wave is partially reflected from and partially transmitted through 
the molecule. In the present work, we admit only infinitesimal voltage differentials, i.e., we 
consider the zero-bias conductance and assume that the ambient temperature is zero. We will be 
concerned with noninteracting spin-polarized contacts which contain up-spin electrons only. This 
requirement can be fulfilled experimentally by using ferromagnetic leads with parallel configurations 
\cite{Seneoretal07JPCM}. Also, external magnets or impurities in the contacts can be employed to 
spin-polarize the contacts \cite{Sachrajdaetal03PRL,Rochaetal05}. The molecule constitutes a subsystem 
in which the electrons repel each other and correlation effects are accounted for by means of a 
multi-configurational description. By constraining the wave function, we ensure that only a single 
electron can escape from the molecule into the contacts. Having two electrons in the contact would 
mean that the wave function in the contact is a multi-electron and not a single-electron one. Apart 
from spin-polarized contacts, our model might be relevant to experiments \cite{CHM06,CHM04}, where hydrogen 
molecules are confined between metal wires. The charging energy of the molecule in such wires might result 
in Coulomb blockade and explain the appearance of just one open channel. While the contacts are non-interacting 
one-electron regions, the interface between the molecule and the contact requires some attention. The 
interface becomes relevant for configurations in which one electron is on the molecule and the other one 
is in the contact. Since there are only up-spin electrons in the contacts, the electron remaining in the 
molecule has down spin. One might think that the electrons are either singlet or triplet coupled, but in fact 
a mixture of singlet and triple state is required to account for the present electron distribution. This is 
because the contact electron is an up-spin electron which cannot be realized with a triplet or a singlet 
coupled configuration, but with an appropriate superposition of the two. This is an important point for 
the basis set selection since in our approach, we calculate the stationary states of a model Hamiltonian 
in finite basis set representation. 

The configuratons that we consider as basis functions are determined such
that they can describe known simple limits. For our purposes, the most important such limit is the dissociated 
molecule where the two atoms of the molecule are decoupled. The incoming up-spin electron in the left contact 
is completely reflected and it does not reach the right contact. Furthermore, due to electron-electron 
repulsion, the up-spin and down-spin electrons are localized in the left and right part of the system 
respectively. Another obvious limit is the non-interacting on. In this case, we have to recover the known 
transmission probability ($T(E)$) curve (see, e.g. \cite{GEZ07}) exhibiting a resonance due to the HOMO and 
one due to the LUMO orbital. One of the key results obtained from our model is that the separation between 
the HOMO and the LUMO resonance is drastically reduced as the electron interaction is introduced. 

We study $T(E)$  of this system as a function of the energy with which the electron travels through the contact, 
i.e., as a function of the Fermi energy at zero temperature. Alternatively, one might think of a gated device in 
which the Fermi energy of the contacts is kept fixed and the molecular part of the MED is shifted in energy by a 
gate voltage. In doing so, we obtain resonances that are related to the molecular states. Therefore, before 
investigating the MED, we consider the isolated molecule (i.e. sans contacts) subject to electron-electron 
interaction. The Hamiltonian of the isolated molecule is given by
\begin{equation}
\sum_{\sigma} \left(\beta_M c^{\dagger}_{a \sigma} c_{b \sigma}+H.c.\right)
+\sum_{\alpha \sigma}h_{\alpha} n_{\alpha \sigma}
+\frac{1}{2} \sum_{\alpha \sigma \sigma'} U n_{\alpha\sigma} n_{\alpha \sigma'}
\label{molham}
\end{equation}
where the operator $c^{\dagger}_{\alpha\sigma}$ ($c_{\alpha\sigma}$) with $\alpha$=a,b
creates(destructs) an electron in the molecular sites a and b.
$\sigma$ stands for the spin degrees of freedom and $n_{\alpha \sigma}$
is the number operator. $U$ is the Hubbard electron-electron repulsion parameter and
$\beta_M$ the intramolecular hopping parameter. The diagonal elements $h_a$ and $h_b$ are 
set equal to zero, thus we will be concerned with a homonuclear molecule.
We consider a finite Hilbert space in which to diagonalize the molecular Hamiltonian.
The singlet configurations that form the basis functions can be divided into two ionic 
ones with both electrons localized on either atom $a$ or atom $b$ and a covalent one with 
one electron per atom, i.e.,
\begin{eqnarray}
\Phi_1 &=& \frac{1}{\sqrt{2}} \mid a a > (\mid  \uparrow \downarrow > - \mid \downarrow \uparrow >),  \nonumber \\
\Phi_2 &=& \frac{1}{2}(\mid a b > + \mid b a >)  (\mid  \uparrow \downarrow > - \mid \downarrow \uparrow > ), \nonumber \\
\Phi_3 &=& \frac{1}{\sqrt{2}} \mid b b >( \mid  \uparrow \downarrow > - \mid \downarrow \uparrow > ).
\label{states}
\end{eqnarray}
This basis enables us to describe the ground state (doubly occupied HOMO) and two 
excited states where one electron or two electron are promoted to the LUMO, respectively.
The Hamiltonian in the finite subspace is then given by \cite{AB02}
\[ \bH= \left( \begin{array}{ccc}
U & \sqrt{2}\beta_{M} & 0  \\
\sqrt{2}\beta_{M} & 0 & \sqrt{2}\beta_{M}  \\
0 & \sqrt{2}\beta_{M} & U
\end{array} \right).\]
Solving the eigenvalue equation for the isolated molecule, we obtain three
eigenvalues,
\begin{eqnarray}
E_0 &=& \frac{1}{2} \left(U-\sqrt{U^2+16 \beta_M^2}\right) \nonumber \\
E_1 &=& U \nonumber \\
E_2 &=& \frac{1}{2} \left(U+\sqrt{U^2+16 \beta_M^2}\right).
\lbl{h2sing}
\end{eqnarray}
As $U$ is increased, the ground state raises in energy to reach zero 
for $ U= \infty$. In the corresponding wave function, there is one electron 
on each atom, completely eliminating the repulsion energy. Due to this 
localization of the electrons there is no binding interaction between 
the atoms either. Furthermore, since our reservoirs are spin polarized, 
the triplet state of the system
\be
\Phi_4 = \frac{1}{2}(\mid a b > - \mid b a >) ( \mid  \uparrow \downarrow > + \mid \downarrow \uparrow > ). \label{trips}  \\
\ee
also plays an important role. The energy of this state is zero, independent 
of the value of $ U $. In the limit where $ \beta_m $ goes to zero or where 
$ U $ goes to infinity, the singlet ground state and the triplet
state become degenerate. For the MED this implies that in these limits, a
superposition of singlet and triplet state determines the transmission 
probability through the system.

We can now turn our attention to the MED, which consists of the desribed diatomic
molecule attached to the left (L) and right (R) spin-polarized semi-infinite contacts 
in serial arrangement. The contacts are modeled by one-dimensional chains of atoms, 
nearest neighbor coupled with strength $ \beta_C $. The diagonal elements of the contact 
atoms are set to zero. L and R in turn, are coupled to atom $ a $ and $ b $, respectively, 
through the parameter $ \beta_{CM} $. Employing the SSP approach \cite{GEZ07} to eliminate 
the infinite contacts, we obtain the Hamiltonian
\begin{equation}
H=h_{\uparrow}+h_{\downarrow}+V_{\uparrow\downarrow}
\label{medham}
\end{equation}
where
\begin{eqnarray}
h_{\uparrow}&=&\Sigma_L n_{L\uparrow}+\Sigma_R n_{R\uparrow}+
\beta_{CM}c^{\dagger}_{L\uparrow}c_{a\uparrow}+H.c.+ \nonumber \\
& & \beta_{CM}c^{\dagger}_{R\uparrow}c_{b\uparrow}+H.c.+\beta_M c^{\dagger}_{a \uparrow}c_{b \uparrow}+H.c.
\end{eqnarray}
\begin{eqnarray}
h_{\downarrow}&=&\beta_{CM}c^{\dagger}_{L\downarrow}c_{a\downarrow}+
H.c.+\beta_{CM}c^{\dagger}_{R\downarrow}c_{b\downarrow}+H.c.+  \nonumber \\
& & \beta_M c^{\dagger}_{a \downarrow}c_{b \downarrow}+H.c.
\end{eqnarray}
and
\begin{equation}
V_{\uparrow\downarrow}=\sum_{\alpha} U n_{\alpha\uparrow}n_{\alpha\downarrow}.
\end{equation}
The operators $c^{\dagger}_{\alpha\sigma}$ ($c_{\alpha\sigma}$) with $\alpha$=$a$, $b$ and 
$c^{\dagger}_{\beta\sigma}$ ($c_{\beta\sigma}$) with $\beta$=L,R create(destruct)
an electron in the molecular sites a and b and contacts L and R respectively.
The source $ \Sigma_L =\beta_C \frac{e^{-iq}+re^{iq}}{1+r} $ and sink $ \Sigma_R =\beta_C e^{iq} $ 
potential \cite{GEZ07} appearing in the Hamiltonian are applied to the first atom of L  and R respectively.
In these potentials, $ q=arccos\left(\frac{E}{2\beta_C}\right) $. They represent the infinite parts of 
single-electron contacts in a rigorous manner \cite{GEZ07}.

\begin{figure}[htb]
\centerline{\includegraphics[angle=0,width=7.5cm]{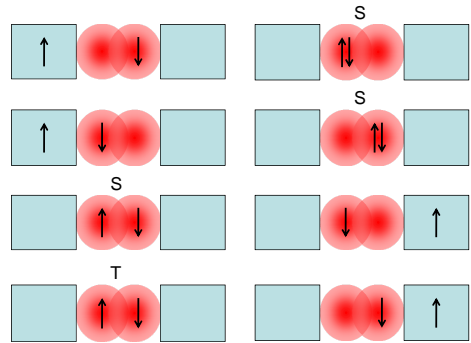}}
\caption{
This figure schematically shows the states used to construct
the Hamiltonian matrix for the full MED. The arrows represent the
spin orientation of the electron and their positions depict where
the electron is localized within the MED. The letters
denote whether the diagram corresponds to a singlet or
triplet.
}
\label{configurations}
\end{figure}

Building on the discussions above, our finite dimensional Hilbert space consists of the for configurations 
listed in Eqs.~\ref{states} and \ref{trips}. We additionally include configurations that are obtained from 
the described ones by exciting the up-spin electron from atom $ a $ or atom $ b $ into the contacts. 
The resulting eight configurations are depicted in Fig.~\ref{configurations}.
%
The Hamiltonian Eq.~\ref{medham} is now represented in the space of the configurations,
\[  \left( \begin{array}{cccccccc}
\Sigma_L & \beta_M  & 0 & 0 & \beta_{CM} & 0 & 0 & 0 \\
\beta_M  & \Sigma_L &  \frac{\beta_{CM}}{\sqrt{2}} & \frac{\beta_{CM}}{\sqrt{2}} & 0 & 0 & 0 & 0 \\
0 & \frac{\beta_{CM}}{\sqrt{2}} & 0 & 0 & \sqrt{2}\beta_{M} & \sqrt{2}\beta_{M} & \frac{\beta_{CM}}{\sqrt{2}} & 0 \\
0 & \frac{\beta_{CM}}{\sqrt{2}} & 0 & 0 & 0 & 0 & -\frac{\beta_{CM}}{\sqrt{2}} & 0 \\
\beta_{CM} & 0 & \sqrt{2}\beta_{M} & 0 & U & 0 & 0 & 0 \\
0 & 0 & \sqrt{2}\beta_{M} & 0 & 0 & U & 0 & \beta_{CM} \\
0 & 0 & \frac{\beta_{CM}}{\sqrt{2}} & -\frac{\beta_{CM}}{\sqrt{2}} & 0 & 0 & \Sigma_R & \beta_M \\
0 & 0 & 0 & 0 & 0 & \beta_{CM} & \beta_M & \Sigma_R \\
\end{array} \right),\]
Having defined our model Hamiltonian, we can proceed to describe how the reflection coefficients are calculated.
The transmission probability $T(\epsilon)$ gives the probability of an electron arriving with an energy $ \epsilon $ 
to pass through the molecule into the right contact. $T(\epsilon)$ is related to the reflection coefficient $r(\epsilon)$ 
by $T(\epsilon)=1 -|r(\epsilon) |^2$. $r(\epsilon)$ is an as of yet unknown parameter in the potential $\Sigma_L$.
To obtain $r(\epsilon)$, we start from the eigenvalue equation
\be
H(\epsilon,r) \Psi = E \Psi,
\label{eigen}
\ee
where $E$ is the total energy of the system. $\epsilon$ takes a real value of our
choice. The total energy of the system can now be written as $E=E^{N-1}+\epsilon $ \cite{KD80,EBP96}, 
where $ E^{N-1} $ is the energy of the $N-1$-electron system obtained by removing the up-spin electron.  
In the present example $E=E^{N-1}$ can take on two different values, one being the orbital energy of 
the binding and the other being the orbital energy of the anti binding orbital. Since we are interested 
in the state that emerges from the molecular ground state, we set $ E=E_b^{N-1}+\epsilon$, where $E_b^{N-1}$ 
is the energy of the binding orbital in the isolated molecule. Therefore, Eq.~\ref{eigen} becomes
\be
H(\epsilon,r) \Psi = ( E_b^{N-1} + \epsilon ) \Psi.
\label{eigen2}
\ee
The corresponding secular equation is
\be
Det (H(\epsilon,r) - (E_b^{N-1} + \epsilon ))=0.
\label{secular}
\ee
Since the value of $ \epsilon  $ represents a  boundary condition that we
specify, Eq.~\ref{secular} determines the unknown $ r $ in terms
of the variable $ \epsilon $.

\begin{figure}[h]
\includegraphics[width=5cm]{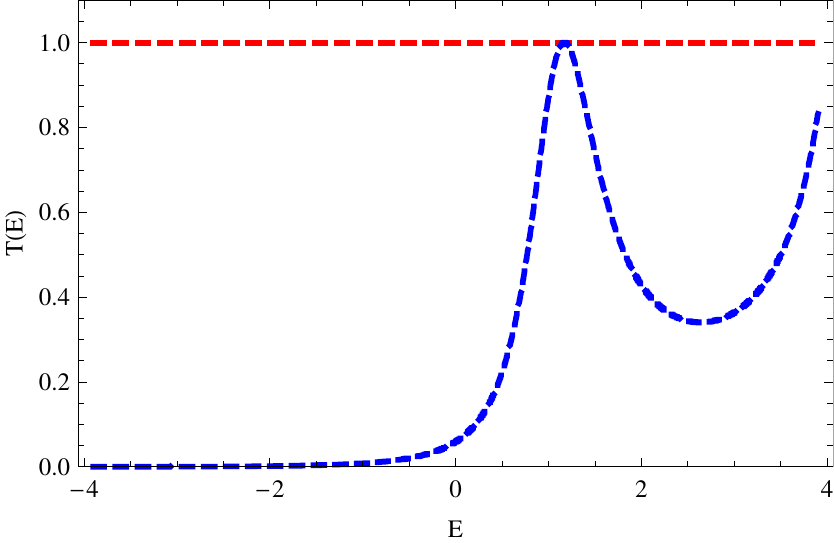}
\caption{This figure shows $T(\epsilon)$ for a homogeneous non-interacting
wire as a function of Fermi energy $\epsilon$. Two solutions are found and 
represented in different colors. The second solution (blue curve) corresponds 
to a wave function that exhibits a localized excited state in the "molecule".}
\label{wire}
\end{figure}

First, we verify that our model reproduces the known answer of $T(\epsilon)=1$
the non-interacting, perfect ballistic conductor. Therefore, we chose a homogeneous linear wire 
with $\beta_C=\beta_{CM}=\beta_{M}=-2$ and $U=0$. Note that we use arbitrary energy units as only 
the relative proportions of the parameters are of significance. Calculation of $T(\epsilon)$ 
according to the above procedure yields the desired solution (red dashed curve in Fig.~\ref{wire}).
We also find a second solution (blue dashed curve) that varies strongly. This curve 
originates from the many-electron basis employed for the central region (supermolecule) 
of the MED comprising of the molecule plus the two contact atoms explicitly considered. 
Even when $U$=0, we still have a many-electron wave function in the supermolecular region. 
This means that we have to expect resonances that correspond to local excitations in the 
supermolecule. Such excited states also couple to the contacts and yield a $T(\epsilon)$ 
that is neither constant nor equal to one. In the localized excited state, the down-spin 
electron is in the anti-bond and the incoming electron passes through the HOMO. Since such 
a local excitation is unlikely to remain localized in a system rather than diffusing into the 
contacts, we think that the second curve is not observable. In our simple treatment, we lack 
the basis functions in the contacts that would permit the local excitation to decay. While this 
excitation by itself might not necessarily be significant, its impact on the non-excited solution 
is. We pursue this issue for inhomogeneous MEDs, where it is realistic to have a strong 
electron-electron interaction in only the molecular part of the MED.

\begin{figure}[h]
\centerline{\includegraphics[width=4.3cm]{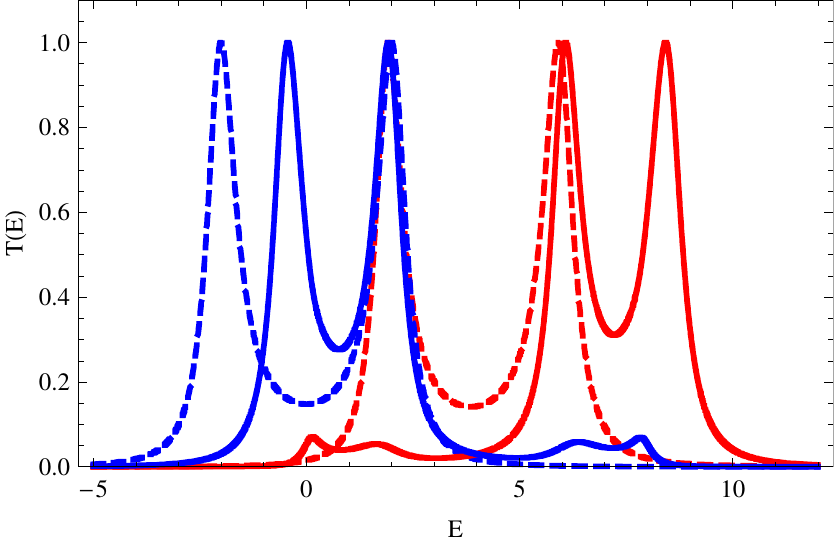}
\includegraphics[width=4.3cm]{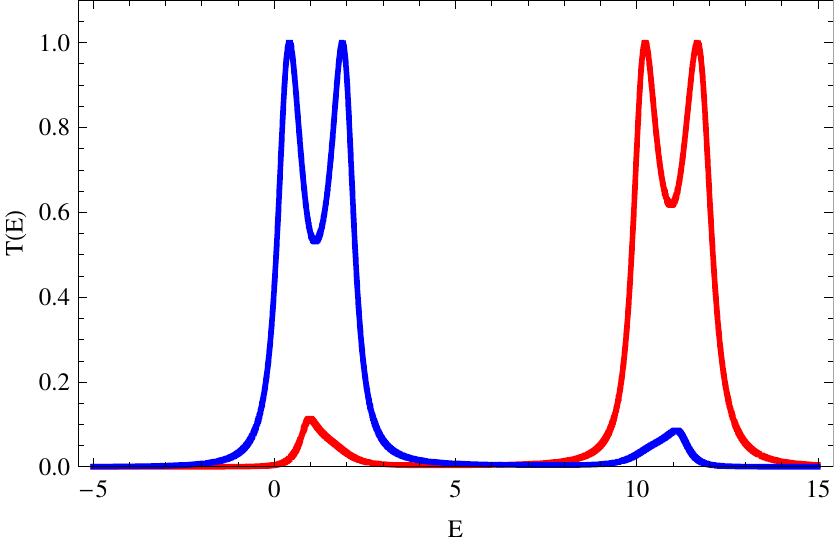}}
\caption{Left panel shows $T(\epsilon)$ for an inhomogeneous wire containing
a diatomic molecule for $U$=0 (dashed curves) and for $U$=4 (solid curves)
as a function of Fermi energy $\epsilon$. Right panel depicts the same system for $U$=8 }
\label{wire-molecule}
\end{figure}

We now consider an inhomogeneous system, where the contact atoms
are more tightly bound $\beta_{C}=-10$ than the remaining ones
($\beta_{CM}=-2,\beta_{M}=-2$). One might think of this system as a stretched break
junction, with a weakened molecule-contact and molecular bond. The
corresponding $T(\epsilon)$ curves for vanishing $U$ are shown in
Fig.~\ref{wire-molecule}. The blue dashed curve on the left, emerging
from the ground state of the molecule has a resonance at the HOMO
energy and a second one at the LUMO energy. The separation of these
orbitals in the isolated molecule is 4($2\beta_M$) and this separation
is essentially preserved \cite{Ernzerhof07} in the MED since the value of
$\beta_{CM}$ is large, this is because, we are close to the wide band limit 
where the contacts become featureless. Note that this curve coincides with the 
one obtained from a non-interacting theory (e.g., SSP as described in Ref.~\cite{Ernzerhof07}).
The red dashed curve, emerging from an excited state of the molecule
shows a resonance at the LUMO energy and a further resonance corresponding
to a doubly occupied LUMO orbital. The red-dashed curve is absent in an
independent electron treatment. Even though the interaction parameter is
$U=0$, we still observe excited states (and the corresponding resonances)
for the reasons discussed above. As in the example of a homogeneous wire, the 
excited-localized state would probably be quenched by the contacts and difficult to 
observe experimentally, unless the coupling between the contacts and the molecule is small. 
For finite $U$, the excited states couples with the ground state and therefore, becomes 
noticeable in the $T(\epsilon) $ curve corresponding to the ground state.

\begin{figure}[h]
\centerline{\includegraphics[width=5cm]{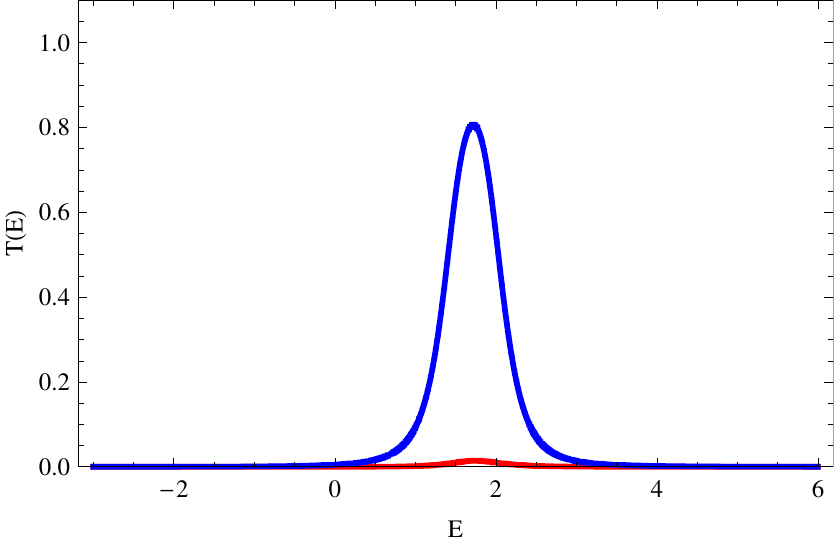}}
\caption{$T(\epsilon)$ for an inhomogeneous wire approaching the
strong correlation limit ($U=32$) as a function of Fermi energy $\epsilon$.}
\label{u32}
\end{figure}

In the interacting system that we are going to consider next we set $ U=4 $.
This choice is motivated by the fact that in the isolated hydrogen molecule,
$ U $ has to be twice as large as $ |\beta_M| $ in order to describe the certain 
properties of the hydrogen molecule \cite{CL07}.
For an interacting system, with $U=4$, the ground
state is destabilized compared to the non-interacting system
(see Eq.~\ref{h2sing}). Furthermore, the separation between the
ground- and the first-excited state increases to ~6.5. Therefore,
naively one would expect a corresponding increase in the separation
between the resonances in the $T(\epsilon)$ curve. We investigate this
question with the help of the solid curves in Fig.~\ref{wire-molecule}
where $U=4$. Surprisingly, the separation between the first two resonances
diminishes even compared to the $U=0$ case. This is due to the
mixing of states that yield the coinciding resonances in the $T(\epsilon)$
curve of the non-interacting case. Furthermore, the second solution
(red solid curve) is shifted towards higher energies due to the
electron-electron repulsion. The states in the red solid curve
have large ionic contributions in which two electrons are localized
on one molecular atom. Increasing $U$ even further to a value of $U=8$
yields the curve depicted in the right panel of Fig.~\ref{wire-molecule}.
The gap between the two resonance diminishes further and the
resonances in the red curve move higher in energy. Eventually, when
the two resonances in the blue curve start to overlap with each other
destructive interference sets in, leading to the disappearance
of $T(\epsilon)$. In Fig.~\ref{u32} we show the transmission results for $U=32$.
In the large $U$ limit, the ionic states described above acquire an infinite
energy due to electron repulsion. The two covalent states, having an electron
on atom $a$ and a second one on atom $b$ in a singlet and triplet
spin-coupling, become degenerate. As a consequence of this degeneracy, we
get a destructive interference between the molecular singlet and triplet state
leading to a localization of up-spin electron and down-spin electron on atom
$a$ or atom $b$ respectively. This interpretation can be verified by
examining the coefficients if the various configurations in the wave function.

Next, we consider the case where the coupling between the molecule and the
contacts is weak $ \beta_{CM} < 1$. We obtain a $T(\epsilon)$ curve that resembles 
the one of Fig.~\ref{wire-molecule}, except, of course, the peak widths are reduced. 
Increasing the interaction parameter($U=4$) yields similar results to the ones shown 
in Fig.~\ref{wire-molecule}. Again, we observe a dramatic reduction of the separation 
between the first two resonances as compared to the non-interacting case.

Now we turn to the dissociation limit, where the coupling between
the molecular atoms ($\beta_M $) approaches zero. In this case, on the scale of the
$ \beta_M $, the importance of the repulsion parameter $U$ increases and we enter the 
strong-correlation regime. Therefore, we expect the same qualitative behavior as in the 
strongly interacting case.

\begin{figure}[h]
\centerline{\includegraphics[width=4.3cm]{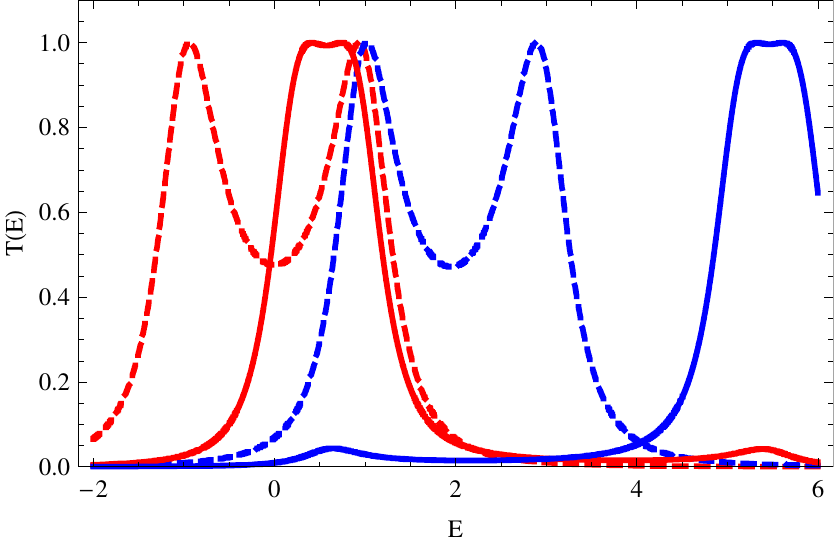}
\includegraphics[width=4.3cm]{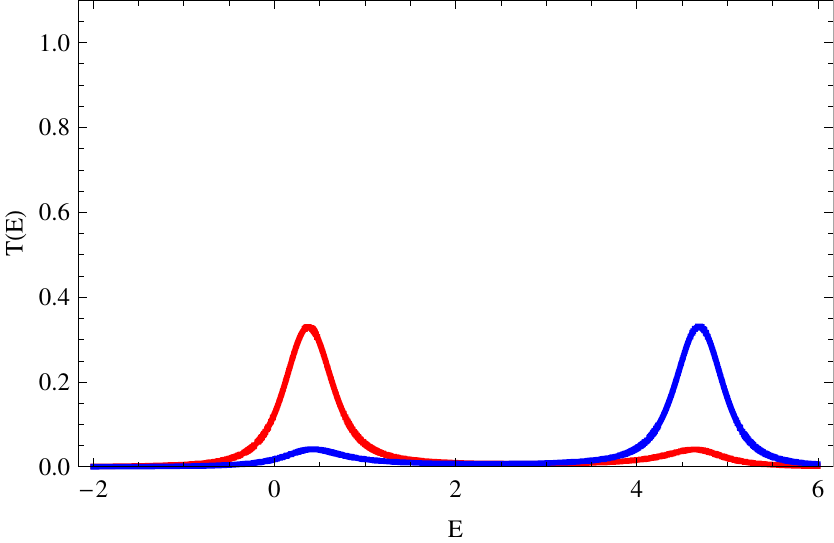}}
\caption{The left plot shows $T(\epsilon)$ for weakly coupled ($\beta_M =-1$) molecular atoms. 
The dashed curves correspond to non-interacting system. The solid curves are obtained for the same 
system with $U=4$. The right plots illustrates  $T(\epsilon)$ for very weakly coupled molecular 
atoms ($\beta_M=-1/2$) with $U=4$.}
\label{weak-bond}
\end{figure}

In the case where $U=0$ and $\beta_M =-1$ (Fig.~\ref{weak-bond}) all the resonances are 
compressed into a smaller energy interval as expected. This implies that electron
interaction is going to have a large effect, since all these close resonances are going 
to be coupled. The solid blue curve, where $U=4$ yields two coinciding resonances, or 
equivalently, the gap between the first two resonances disappeared completely. For cases 
where the Fermi level is located between the HOMO and LUMO level of the non-interacting case, 
a strong conductance enhancement would be obtained. As already explained, this enhancement is 
due to the indirect impact of the molecular triplet state. Reducing the absolute value
of $\beta_M$ even further ($\beta_M=-1/2$) in the right plot of Fig.~\ref{weak-bond}) leads to 
a strong suppression of the $ T $  in the interacting case as opposed to the non-interacting 
case. We are again approaching the strongly-correlated limit and naturally, the $T(\epsilon)$ 
curve resembles the one of Fig.~\ref{u32}. The preceding discussion underlines the statement 
made in the introduction, that bond breaking turns the system into a strongly correlated one. 
Therefore, even if in reality, the value of the parameter $U$ is fixed, stretching the bond 
amounts to an effective increase of the parameter relative to the atom-atom coupling. In general, 
one would expect that electron-electron interaction effects reduce $T(\epsilon)$ \cite{kbe06}. 
This is indeed what we observe, except in cases where the electron repulsion mixes in further 
states as it is the case here with the molecular triplet states.

In conclusion, our understanding of correlation effects on the transmission probability of molecular 
conductors is still quite limited. In this letter, we provided a qualitative model of correlation effects. 
We focused on the molecular part of the MED and considered the simplest possible way of introducing electron 
interaction effects therein. In the interacting case, we uncovered a new, unexpected resonance that 
originates from the lowest triplet state of the isolated molecule. In the presence of electron repulsion, 
this state is coupled to the molecular ground state to drastically reduce the gap between the HOMO and the 
LUMO resonance that would be observed in a non-interacting system.

In our model, the contacts were assumed to be spin polarized. Nonetheless, we argue that our model 
calculations pertain to realistic systems.  Indeed, recent experiments \cite{CHM04,CHM06} in which $H_2$ 
molecules are inserted between gold wires give rise to the speculation that there is only one spin 
channel open. Spin-polarization in the contacts appears to be attainable with experimental techniques 
as well. The fact that the conductance is suppressed gradually in our calculations in the strongly 
correlated limit as a function of $\beta_M$ instead of collapsing abruptly serves as further evidence 
that our results are in line with the experimental data \cite{Smitetal02,CHM06,Schonenebergeretal06NL,Xiaetal08NL}. 
Therefore, we have reason to believe that our model sheds light on electron interaction effects despite 
its seemingly unsophisticated nature.

We gratefully acknowledge financial support provided by NSERC.

\bibliographystyle{jacs}
\bibliography{ref}

\begin{thebibliography}{10}

\bibitem{JGA00}
Joachim, G.; Ginzewski, J.~K.; Aviram, A.
\newblock {\em Nature}
\newblock {\bf 2000}, {\it 408}, 541--548.
\bibitem{N01}
Nitzan, A.
\newblock {\em Annu. Rev. Phys. Chem.}
\newblock {\bf 2001}, {\it 52}, 681--750.
\bibitem{NR03}
Nitzan, A.; Ratner, M.~A.
\newblock {\em Science}
\newblock {\bf 2003}, {\it 300}, 1384--1389.
\bibitem{HR03}
Heath, J.~R.; Ratner, M.~A.
\newblock {\em Phys. Today}
\newblock {\bf 2003}, {\it 56}, 43--49.
\bibitem{CHH07}
Chen, F.; Hihath, J.; Huang, Z.; Li, X.; Tao, N.~J.
\newblock {\em Annu. Rev. Phys. Chem.}
\newblock {\bf 2007}, {\it 58}, 535.
\bibitem{Zuticetal04RMP}
Zutic, I.; Fabian, J.; Sarma, S.~D.
\newblock {\em Rev. Mod. Phys.}
\newblock {\bf 2004}, {\it 76}, 323--410.
\bibitem{Bratkovsky08RPP}
Bratkovsky, A.~M.
\newblock {\em Rep. Prog. Phys.}
\newblock {\bf 2008}, {\it 71}, 026502.
\bibitem{Rochaetal05}
Rocha, A.~R.; Garcia-Suarez, V.~M.; Bailey, S.~W.; Lambert, C.~J.; Ferrer, J.;
  Sanvito, S.
\newblock {\em Nature Materials}
\newblock {\bf 2005}, {\it 4}, 335--339.
\bibitem{Seneoretal07JPCM}
Seneor, P.; Bernard-Mantel, A.; Petroff, F.
\newblock {\em J. Phys.: Condens. Matter}
\newblock {\bf 2007}, {\it 19}, 165222.
\bibitem{MKR94}
Mujica, V.; Kemp, M.; Ratner, M.~A.
\newblock {\em J. Chem. Phys.}
\newblock {\bf 1994}, {\it 101}, 6856--6864.
\bibitem{D95}
Datta, S.
\newblock {\em Electronic Transport in Mesoscopic Systems}; Cambridge
  University Press, 1995.
\bibitem{NDL95}
Lang, N.~D.
\newblock {\em Phys. Rev. B}
\newblock {\bf 1995}, {\it 52}, 5335--5342.
\bibitem{XDR01}
Xue, Y.~Q.; Datta, S.; Ratner, M.~A.
\newblock {\em J. Chem. Phys.}
\newblock {\bf 2001}, {\it 115}, 4292--4299.
\bibitem{TGW01a}
Taylor, J.; Guo, H.; Wang, J.
\newblock {\em Phys. Rev. B}
\newblock {\bf 2001}, {\it 63}, 121104(R).
\bibitem{BMO02}
Brandbyge, M.; Mozos, J.-L.; Ordej{\'{o}}n, P.; Taylor, J.; Stokbro, K.
\newblock {\em Phys. Rev. B}
\newblock {\bf 2002}, {\it 65}, 165401.
\bibitem{KBY04}
Ke, S.-H.; Baranger, H.~U.; Yang, W.
\newblock {\em Phys. Rev. B}
\newblock {\bf 2004}, {\it 70}, 085410.
\bibitem{EZ05}
Ernzerhof, M.; Zhuang, M.
\newblock {\em Int. J. Quantum Chem.}
\newblock {\bf 2005}, {\it 101}, 557--563.
\bibitem{SGP06}
Solomon, G.~C.; Gagliardi, A.; Pecchia, A.; Frauenheim, T.; Di-Carlo, A.;
  Reimers, J.~R.; Hush, N.~S.
\newblock {\em J. Chem. Phys.}
\newblock {\bf 2006}, {\it 125}, 184702.
\bibitem{kbe06}
Koentopp, M.; Burke, K.; Evers, F.
\newblock {\em Phys. Rev. B}
\newblock {\bf 2006}, {\it 73}, 121403.
\bibitem{T08}
Thygesen, K.~S.
\newblock {\em Phys. Rev. Lett.}
\newblock {\bf 2008}, {\it 100}, 166804.
\bibitem{GEZ07}
Goyer, F.; Ernzerhof, M.; Zhuang, M.
\newblock {\em J. Chem. Phys.}
\newblock {\bf 2007}, {\it 126}, 144104.
\bibitem{Ernzerhof07}
Ernzerhof, M.
\newblock {\em J. Chem. Phys.}
\newblock {\bf 2007}, {\it 127}, 204709.
\bibitem{ME85}
Moreland, J.; Ekin, J.~W.
\newblock {\em J. Appl. Phys.}
\newblock {\bf 1985}, {\it 58}, 3888--3895.
\bibitem{RAP96}
van Ruitenbeek, J.~M.; Alavarez, A.; {n}eyro, I.~P.; Grahmann, C.; Joyez, P.;
  Devoret, M.~H.; Esteve, D.; Urbina, C.
\newblock {\em Rev. Sci. Instrum.}
\newblock {\bf 1996}, {\it 67}, 108--111.
\bibitem{PY89}
Parr, R.~G.; Yang, W.
\newblock {\em Density-Functional Theory of Atoms and Molecules}; Oxford
  University Press: Oxford, 1989.
\bibitem{DG90}
Dreizler, R.~M.; Gross, E. K.~U.
\newblock {\em Density Functional Theory}; Springer Verlag: Berlin, 1990.
\bibitem{KMR94b}
Kemp, M.; Mujica, V.; Ratner, M.~A.
\newblock {\em J. Chem. Phys.}
\newblock {\bf 1994}, {\it 101}, 5172--5178.
\bibitem{Tayloretal01PRB}
Taylor, J.; Guo, H.; Wang, J.
\newblock {\em Phys. Rev. B}
\newblock {\bf 2002}, {\it 63}, 245407.
\bibitem{Sachrajdaetal03PRL}
Pioro-Ladriere, M.; Ciorga, M.; Lapointe, J.; Zawadzki, P.; Korkusinski, M.;
  Hawrylak, P.; Sachrajda, A.~S.
\newblock {\em Phys. Rev. Lett.}
\newblock {\bf 2003}, {\it 91}, 026803.
\bibitem{CHM06}
Csonka, S.; Halbritter, A.; Mihaly, G.
\newblock {\em Phys. Rev. B}
\newblock {\bf 2006}, {\it 73}, 075405.
\bibitem{CHM04}
Csonka, S.; Halbritter, A.; Mih\'aly, G.; Shklyarevskii, O.~I.; Speller, S.;
  van Kempen, H.
\newblock {\em Phys. Rev. Lett.}
\newblock {\bf 2004}, {\it 93}, 016802.
\bibitem{AB02}
Alvarez-Fern\'{a}ndez, B.; Blanco, J.~A.
\newblock {\em Eur. J. Phys.}
\newblock {\bf 2002}, {\it 23}, 11--16.
\bibitem{KD80}
Katriel, J.; Davidson, E.~R.
\newblock {\em Proc. Natl. Acad. Sci. USA}
\newblock {\bf 1980}, {\it 77}, 4403.
\bibitem{EBP96}
Ernzerhof, M.; Burke, K.; Perdew, J.~P.
\newblock {\em J. Chem. Phys.}
\newblock {\bf 1996}, {\it 105}, 2798.
\bibitem{CL07}
Chiappe, G.; Louis, E.; SanFabián, E.; Verges, J.~A.
\newblock {\em Phys. Rev. B}
\newblock {\bf 2007}, {\it 75}, 195104--1--195104--6.
\bibitem{Smitetal02}
Smit, R. H.~M.; Noat, Y.; Untiedt, C.; Lang, N.~D.; van Hemert, M.~C.; van
  Ruitenbeek, J.~M.
\newblock {\em Nature}
\newblock {\bf 2002}, {\it 419}, 906--909.
\bibitem{Schonenebergeretal06NL}
Gonzalez, M.~T.; Wu, S.; Huber, R.; van~der Molen, S.~J.; Schonenberger, C.;
  Calame, M.
\newblock {\em Nano Lett.}
\newblock {\bf 2006}, {\it 6}, 2238--2242.
\bibitem{Xiaetal08NL}
Xia, J.~L.; Diez-Perez, I.; Tao, N.~J.
\newblock {\em Nano Lett.}
\newblock {\bf 2008}, {\it 8}, XXX.
\end{thebibliography}
\end{document}